\documentstyle[12pt]{article}
\title{Optimizing entropy relative to a channel or a subalgebra}
\author{Armin Uhlmann}
\date{Institut f\"ur Theoretische Physik,
         Universit\"at Leipzig,\\
         Augustusplatz 10/11, 04109 Leipzig, Germany}
\raggedright
\begin{document}
\maketitle

It is my aim to describe tools, particulary the roof concept, to
handle the {\it entropy of a subalgebra with respect to a state}
as defined by Connes, Narnhofer, and Thirring \cite{CNT87}. How
it works is shown in an example. Proofs are only sketched. The paper
extends common work \cite{BNU96} with F.~Benatti and H.~Narnhofer.

Relying on \cite{Weh78} and \cite{OP93}, I start repeating
definitions. They can live on the framework of unital C$^*$-algebras.
But I restrict myself, up to isomorphy, to finite dimensional
ones, i.~e.~to matrix algebras containing with any matrix its
hermitian conjugate. For short I call such an object an {\it algebra}.
I use the convenient channel terminology \cite{OP93}:
A {\it channel} consists of two algebras, the
{\it input} one, ${\cal B}$, and the {\it output} algebra
${\cal A}$, and a completely positive unital mapping $\alpha$,
the {\it channel mapping}, from the output to the input
system: ${\cal A} \to {\cal B}$ (unital = identity preserving).
The state space of the input algebra is denoted by $\Omega$.
A state, $\omega$, will be identified with its density operator.
$\omega \circ \alpha$ is the pullback
of the state to the output algebra. It is the {\it reduced}
density operator. An {\it ensemble}
\begin{equation} \label{ensemble}
{\cal E} = \{ p_j ; \omega_j \}, \quad \sum p_j = 1, \quad p_k \geq 0
\end{equation}
of ${\cal B}$ is a finite set of states together with
weights. Performing the {\it convex sum}
\begin{equation} \label{csum}
{\cal E}  \, \mapsto \,   \omega := \sum p_j \omega_j
\end{equation}
we get a new state. We refer to (\ref{csum}) as a
{\it convex decomposition of} $\omega$ or, equivalently,
as a {\it Gibbsian mixture} of the states $\omega_k$
with coefficients $p_k$.
(\ref{ensemble}) and (\ref{csum}) are called
{\it short} if no coefficient $p_k$ is zero and all
the $\omega_k$ are mutually different. The {\it length}
is the number of terms in the short decomposition or in the ensemble.

The {\it mutual entropy of a channel} with input ensemble
${\cal E}$ reads
\begin{equation} \label{inf1}
I({\cal E}, \alpha) :=
\sum p_j S( \omega_j \circ \alpha, \omega \circ \alpha )
\end{equation}
$S(.,.)$ stands for relative entropy. Like the latter,
coarse graining implies decreasing of $I$. Ohya defines
\cite{OP93} the {\it entropy of a channel with respect to a state}
by
\begin{equation} \label{centropy}
H_{\omega}(\alpha) := \sup_{\cal E} I({\cal E}, \alpha), \qquad
{\cal E} \mapsto \omega
\end{equation}
The original definition, \cite{CNT87}, appears if $\alpha$ is the
{\it inclusion map} from a unital subalgebra, ${\cal A}$,
into the input algebra.
Then, identifying channel and subalgebra, one writes
$H_{\omega}({\cal A})$ or $H_{\omega}({\cal B}|{\cal A})$
for $H_{\omega}(\alpha)$.  Monotonicity is inherited from
(\ref{inf1}) to (\ref{centropy}). Furthermore, $H_{\omega}$ depends
concavely on $\omega$, (see below), and it is non-negative. Good
reasons to adorn a functional with the word "entropy"!

Set $s(x) = - x \ln x$. Replacing $x$ by a density operator and
performing the canonical trace results in the Gibbs-von Neumann
entropy, also called $S$, but depending on one argument only.
Elementary manipulations show, \cite{BNU96},
\begin{equation} \label{HN2}
H_{\omega}(\alpha) =
 S( \omega \circ \alpha ) - R(\omega, \alpha), \quad
R := \inf_{\cal E} \sum p_j S( \omega_j \circ \alpha ), \quad
{\cal E} \mapsto \omega
\end{equation}
$R$ is the {\it convex hull of the function}
$\omega \to S(\omega \circ \alpha)$ on $\Omega$, see \cite{Roc70}.
The convex hull of any function is a convex function.
Thus (\ref{HN2}) is the sum of two concave functions, and
$H_{\omega}$ is concave on $\Omega$.

To calculate the entropy of a reduced density operator is a
straightforward though often cumbersome task. But to handle
$R$ is difficult. An ensemble (\ref{ensemble}) is called {\it extremal},
iff it consists of pure states only. The set of pure states,
$\Omega^{\rm pure}$, coincides with the extremal
part, $\Omega^{\rm ex}$, of the state space, and it is compact.
Because the entropy functional is concave, it suffices to perform
the inf in (\ref{HN2}) with extremal decompositions only.
As short extremal decomposition is called {\it optimal} iff
\begin{equation} \label{opt}
H_{\omega}(\alpha) = I({\cal E}, \alpha), \quad R =
\sum p_j S(\varrho_j), \quad \{ p_j; \varrho_j \} \mapsto \omega
\end{equation}
and the $\varrho_k$ are pure states. \hfill \\
{\bf Lemma 1.}                       \hfill \\
$H_{\omega}$ and $R(\omega)$ are continuous
on $\Omega$. Every $\omega$ allows for an optimal decomposition
(\ref{opt}). One may require that its pure states generate a simplex.
$\Box$

$R$ is known on the extreme boundary. It is continuous there.
In the real space of Hermitian
matrices we associate to every pure state $\varrho$ the matrix
$\varrho + R(\varrho) \underline{1}$. The set of these matrices
constitutes the compact extreme boundary of its convex hull $\Xi$.
The part of $\Xi$, visible from $\Omega$, is the graph of $R$.
Indeed, the smallest real number $\lambda$ satisfying
$\omega + \lambda \, \underline{1} \in \Xi$ equals $R(\omega)$.
Now the first two assertions can be seen.
The last one follows by Caratheodory's
theorem. $\Box$ \hfill \\
{\bf Lemma 2.} \, Denote by $\Phi_{\omega}^{\rm ex}$ the set
of all pure states $\varrho$ in an optimal decomposition of
$\omega$, and by $\Phi_{\omega}$ its convex hull. \hfill \\
(i) $R$ is affine on $\Phi_{\omega}$. \hfill \\
(ii) If an extremal decomposition of a state $\omega'$ is based on
$\Phi_{\omega}^{\rm ex}$, it is an optimal one. \hfill \\
(iii) $\Phi_{\omega}^{\rm ex}$ and its convex hull, $\Phi_{\omega}$,
are compact. $\Box$ \hfill \\

(ii) is a modification of theorem 1 in \cite{BNU96}. (i) is equivalent
with (ii), and (iii) comes from (i) and the continuity of $R$. $\Box$

Now I extend the notations. Let $F$ be a function on $\Omega$.
A set of extremal points of $\Omega$ is called {\it optimal}
for $F$ iff $F$ is affine on its convex hull.
I call $F$ a {\it roof} if every element $\omega$
is contained in the convex hull of an optimal set. By lemma 2, $R$
is a convex and $-R$ a concave roof. It is an easy
exercise to show: If two convex roofs coincide on the extreme
boundary, they are equal one to another. It results: \hfill \\
{\bf Theorem}  \hfill \\
The entropy of a channel with respect to a state is uniquely
characterized as a functional on the state space of the input
algebra as follows. \hfill \\
(i) $H_{\varrho}(\alpha) = 0$ for pure states $\varrho$. \hfill \\
(ii) $H_{\omega}(\alpha)$ is the sum of $S(\omega \circ \alpha)$
and of a concave roof. $\Box$ \hfill \\
I add without proof another fact, based on lemma 2. \hfill \\
{\bf Lemma 3.}  \hfill \\
Let $H_{\omega} = 0$.
Then $\Phi_{\omega}$ is the face of $\omega$ in $\Omega$, and
every vector belonging to the support of $\omega$ is a common
eigenvector for all operators in the output algebra.
$\Box$ \hfill \\

Now I treat two examples to see the roof concept working.
In the first, known one \cite{BNU96},
the input algebra consists of the 2-by-2-matrices. The subalgebra
of its diagonal matrices is the output algebra. From a density
operator $\omega$ we need the off-diagonal entry $z = z_{12}$.
Assume $F(\omega) = f(|z|)$. Such a function is convex on
$\Omega$ iff $f$ depends convexly on $|z|$. Next, the set of
density operators with fixed $z$ is convexly generated
by its pure states. Hence, $F$ is certainly a roof.
From all that we conclude
\begin{equation} \label{m2-3}
R(\omega) = s(q) + s(1 - q), \quad
q :=  {1 \over 2} + {1 \over 2} \sqrt{1 - 4 z \bar z}
\end{equation}
Indeed, equality in (\ref{m2-3}) is true for pure states.
Being of the form $R(\omega) = f(|z|)$ it is a roof. It
remains to see convexity on $|z| \leq 1/2$. Taylor $q$-expanding
(\ref{m2-3}) shows convexity term by term:
\begin{equation} \label{m2-4}
R(\omega) = r_2(z) :=
\ln 2 - \sum_{k=1}^{\infty} {(1 - 4 z \bar z)^k \over 2k (2k - 1)},
\end{equation}
%%%%%%%%%%%%%%%%%%%%%%%%%%%%%%%%
My next example reads
\begin{equation} \label{m2-def}
{\cal B} := {\cal M}_{n+1}, \qquad
{\cal A} := {\cal M}_n \oplus {\cal M}_1
\end{equation}
There are projection operators, $P$ and $Q$, in our
input algebra satisfying $P + Q = \underline{1}$,
$P = | \psi \rangle \langle \psi |$, such that
the reduced density operator and its entropy is gained by
\begin{equation} \label{m2-6}
\omega \circ \alpha = Q \omega Q + \lambda P, \quad
\lambda = \langle \psi, \omega \, \psi \rangle
\end{equation}
\begin{equation} \label{m2-7}
S(\omega \circ \alpha) =  s(\lambda) +
{\rm Tr} \, s(Q \omega Q)
\end{equation}
I like to compute $R$ and to describe $\Phi_{\omega}^{\rm ex}$.
We choose orthonormal eigenvectors $\psi_1, \dots, \psi_n$
of $Q \omega Q$ such that
\begin{equation} \label{m2-7x}
z_k := \langle \psi_k, \omega \, \psi \rangle \geq 0 , \quad
\lambda_k = \langle \psi_k, \omega \, \psi_k \rangle
\end{equation}
Then $\lambda_j \lambda \leq z_j^2$.
Trying to find an ansatz for optimal sets I define
\begin{equation} \label{m2-8x}
z_j = (p_j^{+} + p_j^{-}) z, \quad \lambda_j =
 p_j^{+} \mu^{+} + p_j^{-} \mu_j^{-}
\end{equation}
so that
\begin{equation} \label{m2-9x}
\varrho^{\pm}_k = z \, |\psi_k \rangle \langle \psi | +
z \, |\psi \rangle \langle \psi_k | +
\mu^{+} |\psi_k \rangle \langle \psi_k| +
\mu^{-} |\psi \rangle \langle \psi|
\end{equation}
defines pure states. This is possible with
\begin{equation} \label{mneu-1}
z = \sum |z_j| \leq {1 \over2}, \quad \mu^{\pm} =
{1 \pm \sqrt{1 - 4 z^2} \over 2}
\end{equation}
and in that case
\begin{equation} \label{m2-9a}
\omega = \sum  p_j \varrho^{+}_j + (1 - p_j) \varrho^{-}_j
\end{equation}
is an essentially unique extremal convex decomposition of $\omega$.
We get
\begin{equation} \label{mneu-2}
R(\omega) \leq s(\mu^{+}) + s(\mu^{-}) = r_2(\sum |z_j|)
\end{equation}
Can the equality sign be true and can (\ref{m2-9a}) be optimal?
As long $z \leq 1/2$ is fulfilled, (\ref{mneu-2}) defines a
roof that coincides for pure states with (\ref{m2-7}).
(\ref{mneu-2}) is invariant with respect to unitaries
from ${\cal A}$, and the set of all $Q \omega Q$ is a unitarily
invariant convex set of Hermitian $n \times n$-matrices.
Such a functional is convex if its restriction
to the diagonal matrices in $Q \Omega Q$ is convex \cite{AU81}.
We obtained a convex roof on the considered part of $\Omega$.
This looks hopefully. However, it remains the question, whether
the other part, $z > 1/2$, of $\Omega$ can beat it by bifurcating
the roof to another one. \hfill \\
%%%%%%%%%%%%%%%%%%%%%%%%%%%%%%%
{\bf Appendix} : {\it Accessible information} \hfill \\
A channel $\alpha$ is called a {\it communication channel} iff
the output algebra is commutative and of finite dimension.
Using the expression (\ref{inf1}) for mutual entropy one defines
\begin{equation} \label{a-1}
I({\cal E}) := \sup_{\alpha} I({\cal E}, \alpha)
\end{equation}
where the sup runs through all communication channels. \hfill \\
Now let ${\cal C}$ be a unital commutative $^*$-subalgebra and
$\omega$ a density operator of the input algebra ${\cal M}_n$.
${\cal C}$ is the linear span of an orthogonal set
$\{ Q_1, \dots, Q_r \}$ of projection operators which sum up to
the identity. Together with $\omega$ they determinate an ensemble
\begin{equation} \label{a-2}
{\cal E} := \{ \, p_j, \varrho_j \, \}, \quad
p_k = {\rm Tr} \, Q_j \omega, \quad p_j \varrho_j =
\sqrt{\omega} Q_j \sqrt{\omega}
\end{equation}
Within this setting Benatti \cite{Ben96} has shown
\begin{equation} \label{a-3}
I({\cal E}) = H_{\omega}({\cal C})
\end{equation}
This helps in computing $I$ and in understanding the observed
similarities in the behaviour of these quite different concepts.
See \cite{Lev94}, \cite{FC94}, \cite{Fu95}, \cite{FP96}. \hfill \\
Observe that the Holevo bound,
\cite{Ho73}, can be seen from (\ref{a-3}) by the
monotonicity of $H_{\omega}$.
\begin{equation} \label{a-4}
H_{\omega}({\cal C}) \leq H_{\omega}({\cal M}_n) \equiv S(\omega)
\end{equation}

$$  $$
Abstract: \hfill \\
After recalling definition, monotonicity, concavity, and continuity
of a channel's entropy with respect to a state (finite dimensional
cases only), I introduce the roof property, a convex analytic tool,
and show its use in treating an example. Full proofs and more examples
will appear elsewhere. The relation (a la Benatti) to accessible 
information is mentioned.          \hfill \\
To be published in: Proceedings of the XXI International
Colloquium on Group Theoretical Methods in Physics, Goslar 1996
$$  $$
{\it e-mail address} : \, \,
Uhlmann@tph100.physik.uni-leipzig.de

\end{document}